\documentclass{article}
\usepackage[margin=1.0in]{geometry}                
\usepackage{geometry}
\usepackage{euscript,amsmath,amssymb,amsfonts,graphicx,bm}
\usepackage{hyperref}
\hypersetup{colorlinks=true,linkcolor=blue,citecolor=red}
\usepackage{epstopdf}
\usepackage{enumerate}
\usepackage{csquotes}
\usepackage{caption}
\usepackage{subcaption}
\usepackage{mathrsfs}
\usepackage{xcolor}

\usepackage{lineno}

\usepackage{mathtools} 
\usepackage{amsthm}
\usepackage{authblk}

\makeatletter
\renewcommand{\maketitle}{\bgroup\setlength{\parindent}{0pt}
\begin{flushleft}
  \textbf{\LARGE \@title}

  \@author
\end{flushleft}\egroup
}
\makeatother
\author[1]{Ethan Levien}
\author[2]{Jiseon Min}
\author[3]{Jane Kondev}
\author[1]{Ariel Amir}
\affil[1]{John A. Paulson School of Engineering and Applied Sciences, Harvard University, Cambridge, MA 02138, USA }
\affil[2]{Department of Molecular and Cellular Biology, Harvard University, Cambridge, MA 02138, USA}
\affil[3]{Department of Physics, Brandeis University, Waltham, MA 02453 USA}

\makeatletter
\patchcmd{\@maketitle}{\LARGE \@title}{\fontsize{20}{19.2}\selectfont\@title}{}{}
\makeatother

\title{Non-genetic variability: survival strategy or nuisance?}
\date{\today}
\begin{document}

\maketitle
\section*{Abstract}
The observation that phenotypic variability is ubiquitous in isogenic populations has led to a multitude of experimental and theoretical studies seeking to probe the causes and consequences of this variability.  Whether it be in the context of antibiotic treatments or exponential growth in constant environments, non-genetic variability has shown to have significant effects on population dynamics. 
Here, we review research that elucidates the relationship between cell-to-cell variability and population dynamics. After summarizing the relevant experimental observations, we discuss models of bet-hedging and phenotypic switching. In the context of these models, we discuss how switching between phenotypes at the single-cell level can help populations survive in uncertain environments.  Next, we review more fine-grained models of phenotypic variability where the relationship between single-cell growth rates, generation times and cell sizes is explicitly considered. Variability in these traits can have significant effects on the population dynamics, even in a constant environment. We show how these effects can be highly sensitive to the underlying model assumptions.  We close by discussing a number of open questions, such as how environmental and intrinsic variability interact and what the role of non-genetic variability in evolutionary dynamics is. 
\newpage
\section{Motivation and background}
\begin{figure}[h!]
\centering
\includegraphics[scale=0.7]{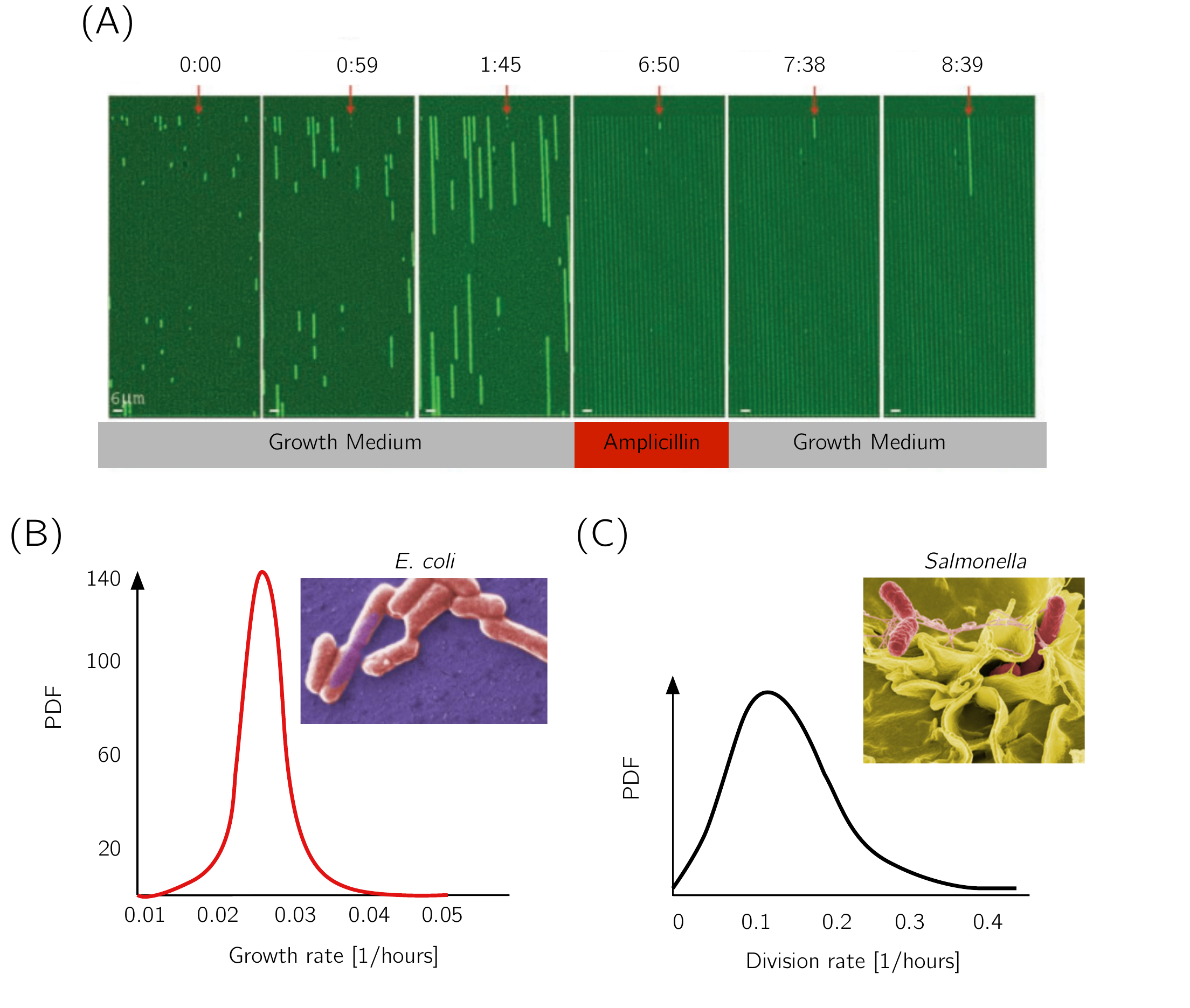}
\caption{(A) The microfluidic device used to detect persister cells in ref.  \cite{balaban2004}. The green lines are lineages of \emph{E. coli} and the red arrows indicate persister cells. The persisters are shown to be present prior to the treatment with ampicillin. Adapted from ref.  \cite{balaban2004}.  (B) A histogram of growth rates (defined as the derivative of volume normalized by the current volume, averaged over the cell-cycle)  from a colony of  \emph{E. coli}. Adapted from ref. \cite{lin2017} with image from Wikimedia Commons. (C) A histogram of \emph{Salmonella} division rates extracted from an infected mouse. (The division rate is the growth rate multiplied by $\ln(2)$). The distribution is significantly wider than for \emph{E. coli} grown in the laboratory setting, despite the fact that the two bacteria are physiologically similar. Adapted from ref.  \cite{claudi2014} with image from Wikimedia Commons. }\label{fig:1}
\end{figure}

 It has been known that variation among genetically identical cells is present and plays an important role  in microbial populations since the 1940s, when J.W. Bigger first discovered \emph{persister cells}  \cite{bigger1944}. In his seminal experiment, Bigger treated \emph{Staphylococcus pyogenes} cells with penicillin, observing that approximately one millionth of the population, the persisters, survived the antibiotic treatment. In order to reveal how these cells first emerged in the population, Balaban et al. grew \emph{Escherichia coli} in microfluidic channels, allowing them to trace individual lineages of cells  \cite{balaban2004}. After exposing cells to the antibiotic ampicillin, so that only the persisters remained, they traced the lineages back through time and found the persisters emerged stochastically and prior to the antibiotic treatment; see Figure \ref{fig:1} (A).  Because any cell in a growing population has a small chance to stochastically switch into a persister cell, some fraction of the population will always be in the persister state, and as soon as the population is exposed to antibiotics, only the persisters cells remain. Persisters can also emerge when cells are transferred to a new media where they do not immediately grow. Some cells will start growing shortly after the transfer, while the persisters remain dormant for an extended period of time \cite{fridman2014}.  It is now believed that many other pathogens, as well as cancer cells, utilize some form of persistence to maintain slow growing subpopulations which can impede treatments.  
  
Persister cells are merely one of many sources of  \emph{non-genetic variability} in microbial populations. Another well-studied example of a phenotypic switching mechanism is the lactose uptake network in \emph{E. coli}, which has been shown to exhibit bistable dynamics. Combined with stochastic gene expression, these dynamics can result in a bimodal distribution of protein levels. By maintaining a subpopulation which is not optimized to metabolize lactose, it is believed that \emph{E. coli} can hedge their bets against nutrient shifts \cite{mettetal2006,wolf2005}. But phenotypic variability does not always manifest as distinct subpopulations with large physiological differences between them. In recent decades,  advances in single-cell measurement techniques have allowed researchers to quantify variability in nearly every measurable trait (e.g. growth rates, cell sizes, doubling times). In some cases, cell-to-cell variability is strikingly small. For example, in \emph{E. coli} the coefficient of variation (CV) of growth rates (defined as the the standard deviation over the mean) in a growing population is between $.06$ and $0.18$ \cite{godin2010,lin2017,soifer2016,Jafarpour2019}; see Figure \ref{fig:1} (B).  This small CV might suggest growth rate variation is an inevitable byproduct of stochastic processes inside cells, which evolution has found ways to repress.  However, in vivo measurements from mice inoculated with \emph{Salmonella}, another rod shaped, gram-negative bacteria, have revealed extensive variation in growth rates, with the CV being as large as $0.6$ \cite{claudi2014}; see Figure \ref{fig:1} (C). Whether the increased variation is due to intrinsic factors within the cells or environmental factors  is currently unknown.  

In many cases, phenotypic variation is a result of stochastic gene expression, i.e. variation from cell to cell due to small number fluctuations in protein or mRNA concentrations \cite{elowitz2002}. However, the mechanisms for generating phenotypic variability extend beyond stochastic gene expression. By dividing asymmetrically the pathogen \emph{Mycobacteria tuberculosis} can generate progenies which are heterogeneous in size and growth rates \cite{aldridge2012,bergmiller2017}.  Even organisms which divide symmetrically, such as \emph{E. coli}, exhibit small amounts of asymmetry when, for example, damaged proteins accumulate in one of the daughter cells, leading to the deterioration, or \emph{aging} of certain lineages which accumulate the older proteins  \cite{chao2016,stewart2005}.

As experiments have exposed the sources of non-genetic diversity in microbial populations it has become common to ask: when is variability a nuisance or a feature that has been selected for by evolution?    
The purpose of the present review is to survey approaches to modeling the population-level effects of phenotypic variability.  Roughly speaking, there are two broad classes models we will consider: models of microbial population growth with changing environments and models where the environment is kept constant. A classical example of how variability can benefit a population in a changing environment is bacterial  persistence. It appears that the persister cells act as a form of insurance against the rare but catastrophic event of an antibiotic attack.  In this way, a population that is periodically exposed to antibiotics will have a higher fitness in the long-term if there are some persister cells present in the population, even though a population with no persisters usually grows faster. Mathematical models have been used extensively to quantify the precise fitness benefits that are conferred by switching between two (or more) phenotypic states, shedding light on how evolution has shaped this phenomenon \cite{kussell2005,mora2013,thattai2004}.

In theoretical investigations of phenotypic variability in variable environments, it is common to adopt coarse-grained models where the underlying dynamics of cell-growth are not specified. In these models, each cell's phenotypic state is characterized by a single quantity (typically the division rate) and variability is introduced by allowing cells to randomly change their state. In reality, a cell's phenotypic state is determined by numerous variables, each undergoing their own complex, dynamic evolution as the cell grows and divides. Which of these variables are relevant to the population-level dynamics is often unclear. This has motivated studies of more fine-grained models of phenotypic variability, which are typically carried out in constant environments for simplicity. These models explicitly account for the stochastic dynamics of single-cell biomass accumulation and cell division,  and they have revealed important subtleties and challenges associated with understanding cell-to-cell variability. 
    Part of our objective in writing this review is to consolidate research on coarse-grained models in variable environments, and fine-grained models in constant environments.    In doing so, we hope to motivate future progress towards a more unified understanding of cell-to-cell variability and its consequences. A key theme, which we will return to throughout the review, is how model details influence the end conclusions about the fitness effects of variability.   Consider the asymmetric segregation of damaging proteins. Intuitively, partitioning all the proteins into one daughter cell benefits the population by concentrating the damage to a single lineage, thereby allowing the undamaged cells to grows at their maximum potential. On the other hand, perhaps maximum growth is achieved by spreading out the damage equally among all cells. Surely the answer depends on how the growth of cells changes with the amount of damage, but exactly how? Understanding why some organisms divide nearly symmetrically while others appear to deliberately produce offspring of different sizes and biochemical compositions has proven to be a challenging mathematical problem \cite{lin2019optimal,chao2016,ackermann2007}, and similar challenges emerge in other examples.  Finally, we note that this review will not cover the molecular mechanisms that generate cell-to cell variability, as these have been discussed extensively in previous reviews (e.g., ref. \cite{eldar2010}).


\section{The benefits of phenotypic variability in changing environments}
\subsection{The theory of bet-hedging}\label{sec:kelly}
\begin{figure}[h!]
\centering
\includegraphics[scale=0.8]{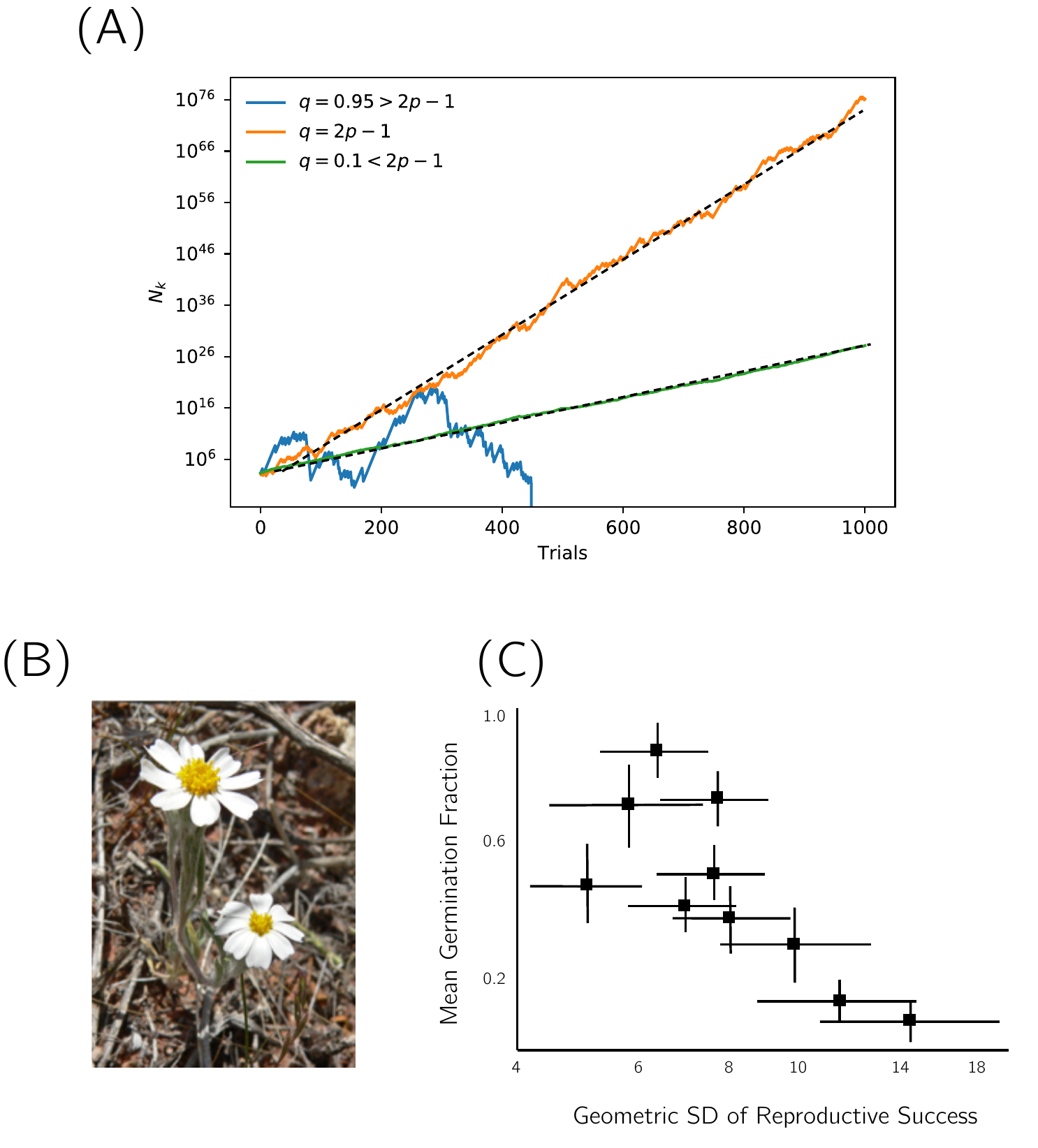}
\caption{(A) Simulations of Kelly's gambling model for different values of $q$ with $p = 0.6$. For values of $q$ which are not too large, we see that the profits approach exponential growth (dashed lines).  (B)  \emph{Eriophyllum lanosum}, a species of wildflower prominent in the southwestern United States, is one of many plants believed to implement a bet-hedging strategy. Image from Wikimedia Commons (C) An inverse relationship between the germination fraction (defined and the fraction of seeds that germinate, or grow into plants, in the season they are dispersed) and the standard deviation in reproductive success of desert annuls  \cite{venable2007}. The latter quantity is the standard deviation of the number of successful plants resulting from seeds that germinate, and thus quantifies the susceptibility of a plant to environmental fluctuations.  This suggests that plants which are more susceptible to environmental fluctuations have evolved to delay germination. Adapted from ref. \cite{venable2007}.   }\label{fig:2}
\end{figure}

It is believed that many sources of cell-to-cell variability are examples of \emph{bet-hedging}. We introduce a technical definition below, but roughly speaking, bet-hedging occurs when organisms trade fitness in the present environment for increased fitness in the event of an environmental transition, such as an exposure to antibiotics.  Originally developed by Kelly in 1956 in the context of gambling \cite{kelly1956}, the idea of bet-hedging is perhaps best illustrated with the following example.  Suppose we are betting on a series of horse races, each involving only two horses. In each race, we may choose to bet a fraction $q$ of our money. Let us assume that the returns from each horse are equal; that is, if we win, the money we have gambled doubles and if we lose all the money we have bet is lost. If the probability of a win is $p > 1/2$, how much of our money should we put on each horse?  The answer depends on our objective. If we wish to optimize the expected value of our returns after a single trial, denoted $N_1$, then we should maximize 
\begin{equation}\label{intro:kelly:N1}
\langle N_1 \rangle  = ((1-q)+2 qp)N_0,
\end{equation}
where $\langle \cdot \rangle$ represents the average over many realizations of the race and $N_0$ is our initial capital. 
The average returns are maximized when we bet all our money ($q=1$), and after $k$ runs the expected value will be
\begin{equation}
\langle N_k \rangle  = (2p)^kN_0.
\end{equation}
However, if this strategy is continued indefinitely we will become broke with probability one because this large expectation comes from the unlikely event of winning $k$ times in a row.  Now consider a scenario where at every trial we gamble a fraction of our money $0 < q < 1$. Equation \eqref{intro:kelly:N1} implies that after $k$ trials, our returns are given by
\begin{equation}
N_k = (1-q)^L(1+q)^WN_0 = N_0e^{L\ln (1-q) + W\ln (1+q)},
\end{equation}
where $L$ and $W$ are the number of losses and wins respectively and are thus subject to the constraint $L+W=k$. For large $k$, the exponent measures the per generation rate at which our profits increase, so we define 
\begin{equation}\label{kelly:Lambda}
 \Lambda = \lim_{k \to \infty} \frac{1}{k}\ln N_k = p\ln(1+q) + (1-p)\ln(1-q).
\end{equation}
Here we have made use of the fact that $p = W/k$ and $1-p=L/k$.
If we intend to continue making a profit while gambling indefinitely,  it is in our best interest to make $\Lambda$ as large as possible. 
To understand why, it is useful to consider the distribution of $N_k$. This can be approximated by noting that $L$ and $W$ are the sum of independent and identically distributed Bernoulli variables, and they can therefore be approximated by Gaussian random variables. In particular $W$ will approach a Gaussian with mean $kp$ and variance $\sigma_W^2 = kp(1-p)$. Since $L = k-W$, we have
\begin{equation}
\ln \frac{N_k}{N_0} = (k-W)\ln (1-q) + W\ln (1+q) = k\ln (1-q) + W \ln \left(\frac{1+q}{1-q}\right). 
\end{equation}
For large $k$, $\ln N_k/N_0$ is a Gaussian with mean $\mu = k \Lambda $ and variance $\sigma^2 =  kp(1-p)(\ln (1+q)/(1-q))^2$. We can now evaluate the mean and variance of $N_k$ to find that 
 the CV will decay exponentially at a rate $\sigma^2$. Since this will vanish for large $k$, the logarithm of our profits will become deterministic around $\Lambda$!
Returning to Equation \eqref{kelly:Lambda}, for $p>1/2$ the maximum value of $\Lambda$ is obtained when $q = 2p-1$. 
Simulations of this model are shown in Figure \ref{fig:2} (A).

One of Kelly's key observations is that for a multiplicative growth process, the long-term outcome is determined not by the average returns, but by the average log returns. This idea was imported into two independent pieces of biological literature \cite{lewontin1969, gillespie1974}. Here, we focus on the work of Cohen and Lewontin, who considered a population growing in a fluctuating environment  \cite{lewontin1969}. Letting $N_k$ denote the size of the population after the $k$-th generation, they supposed that the size at the $(k+1)$th generation is given by 
\begin{equation}\label{Nk_cohen}
N_{k+1} = f_k N_k,
\end{equation} 
where $f_k$ is the number offspring that survive to the $(k+1)$th generation. It is assumed that $f_k$ is drawn independently from some distribution $\rho$ at each generation. This distribution may depend on both properties of the individual cells and the environment; for example, random fluctuations in weather could cause $f_k$ to vary from season to season for a certain species of plants.  At the same time, seeds may vary in their tolerance within a season, which will also affect the distribution of $f_k$ \cite{menu2000,cohen1970}.  The question Cohen and Lewontin pose is: how does the distribution of $f_k$, and in particular the variation in $f_k$, affect the long-term growth of the population? We already know from Kelly's gambling example that maximizing $\langle f_k \rangle$ is a disastrous  strategy from the perspective of the population, as this could lead to the population going extinct with probability of one. To see how variability affects the long term growth of $N_k$, consider the recursive equation of the log population size: 
\begin{equation}
\ln N_{k+1} =\ln  f_k  + \ln N_k \approx \langle \ln  f_k \rangle  + \ln N_k + \xi_k,
\end{equation}
where $\xi_k$ is a noise term with zero mean. This has the form of a random walk in $\ln N_k$ with a drift term given by $\langle \ln f_k \rangle$, which determines the fate of the population in the long-term. Unless this is positive, the population will surely go extinct in the long run. After Taylor expanding $f_k$ around $\langle f_k \rangle$, taking the natural log and averaging, we have 
\begin{equation}\label{bh}
\langle \ln f_k \rangle \approx  \ln \langle f_k \rangle - \frac{1}{2}\frac{{\rm var}(f_k)}{\langle f_k \rangle^2}. 
\end{equation}

 In Kelly's gambling example, the optimal betting strategy reduces the variation in our profits while also reducing the expected profits per trial. In terms of Equation \eqref{bh}, this can be understood as a trade-off between maximizing $\langle f_k \rangle$, known as the \emph{arithmetic mean fitness}, and minimizing ${\rm var}(f_k)$ in order to maximize $\langle \ln f_k \rangle$, the \emph{geometric mean fitness}. As we saw in Kelly's example, rare events contribute disproportionately to the arithmetic fitness, which is why it is the geometric fitness  which determines the long-term growth rate of a multiplicative process.  This leads to the formal definition of bet-hedging: \emph{bet-hedging is a reduction in arithmetic fitness in exchange for geometric fitness} \cite{starrfelt2012}. 
 Much like gamblers hedge their bets, biological populations can diversify their phenotypes to reduce the effects uncertain environmental conditions have on $f_k$, thereby reducing ${\rm var}(f_k)$, often at the expense of  $\langle f_k \rangle$. This strategy is found in many annul plant species, where some fraction of the seeds may not germinate in the same year they were dispersed. This has the effect of reducing the expected number of offspring per year but provides some security against conditions which may kill off any plants that germinate (e.g. droughts). By analyzing data obtained over a 22 year period in Arizona, Venable found an inverse relationship between the germination fraction of desert annuls and their susceptibility to environmental variation \cite{venable2007}; see Figure \ref{fig:2} (B,C).  Venable concluded that plants are implementing the trade-off described by Equation \eqref{bh}.  
\subsection{Phenotypic switching}\label{sec:switch}

\begin{figure}[h!]
\centering
\includegraphics[scale=0.8]{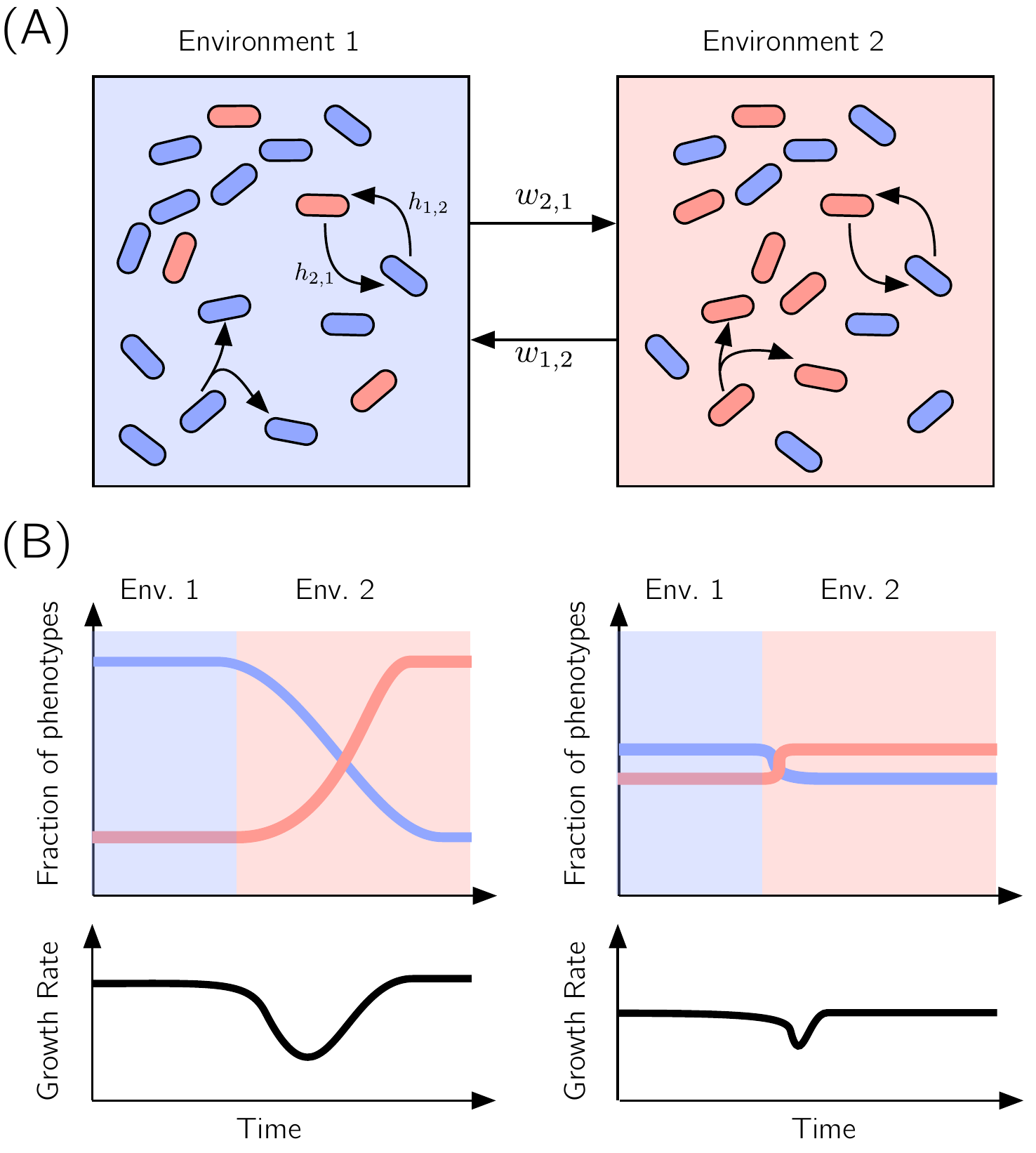}
\caption{(A) A diagram of the phenotypic switching model with 2 environments and 2 phenotypes. In environment 1 phenotype 1 (the blue cells) grow faster, while in environment 2 phenotype 2 (the red cells) grow faster. (B) and (C) Cartoons of trajectories of the fraction of each phenotype and fitness under two different conditions. Here, the $h_{i,j}$ are the rates to switch between phenotypes, while the $w_{i,j}$ are the rates to switch between different environments. (B) On the left the cells switch slowly and therefore the faster growing cells are under strong selection leading to a large instantaneous population growth rate after the population has been in environment 1 for a long period of time. However, when the environment switches it takes awhile for the population to relax to the new steady state. The population plays a price for this slow relaxation during the transient. On the right we see the opposite situation, where phenotypic switching is fast so that the less fit phenotype is in greater abundance in both environments. Here the population growth rate is smaller when the phenotype distribution is in steady-state.  }\label{fig:3}
\end{figure}

As noted in the introduction,  one way in which microbes can generate cell-to-cell variability is through phenotypic switching; that is, through individual cells stochastically changing their phenotypic states in order to optimize their fitness in changing environments.  In order to understand this phenomenon, it is more natural to use a continuous model than the discrete time models formulated by Lewontin  and Gillespie.   The continuous  time setting is often more realistic for microbial populations, because, unlike plants, microbe divides asynchronously. That is, there is no notion of a single generation time for the entire population, rather each individual lives for a random amount of time so that cells are continuously dividing and being born. A useful approximation is to assume that cells divide at a constant rate (in reality this is far from true, but we will postpone a discussion of this point until Section \ref{sec:age_struct}). If every cell in the population is dividing at the same rate $\Lambda$ (meaning $\Lambda dt$ is the probability for a given cell to divide in a small time interval $dt$), the population will grow exponentially at a rate $\Lambda$.  Now consider a population with two phenotypes which grow at rates $\Lambda_1$ and $\Lambda_2$ respectively and suppose that cells of phenotype $k$ switch to phenotype $j$ at a rate $h_{j,k}>0$; see Figure \ref{fig:3} (A). The number of cells in the two phenotypes, denoted $N_1$ and $N_2$, will obey the linear ordinary differential equation: 
\begin{align}\label{Nk_switching}
\frac{d}{dt}\left[\begin{array}{c} N_1\\ N_2 \end{array}\right] &=
\left( \underbrace{\left[\begin{array}{cc} 
\Lambda_1 & 0\\
0 & \Lambda_2
 \end{array}\right] }_{:= G}
 + 
\underbrace{\left[\begin{array}{cc} 
-h_{2,1} & h_{1,2}\\
h_{2,1} & -h_{1,2}
 \end{array}\right] }_{:= H}\right)
 \left[\begin{array}{c} N_1\\ N_2 \end{array}\right]. 
\end{align}
Here we have decomposed the matrix on the right hand side in terms of the growth rate matrix, $G$, and the switching matrix, $H$. 
Assuming the population evolves in a constant environment, $N_{\rm tot} =N_1 + N_2$ will grow exponentially in the long-term with a growth exponent given by the top eigenvalue of $A= G + H$, denoted $\lambda_0$, while distribution of the phenotypes will eventually converge to the corresponding normalized eigenvector, denoted ${\bf q}_0$. Computing formulas for these quantities is an exercise in elementary linear algebra, which we leave to the reader. It is illuminating to consider two limits: When the growth rates are much greater than the switching rates, the top eigenvalue will be approximated by ${\rm max}(\Lambda_1,\Lambda_2)$ and the top eigenvector will approach a standard basis vector, $(0,1)^T$ or $(1,0)^T$, around the faster growing phenotype. In the opposite limit, when cells switch between states very rapidly relative to the growth rate, the distribution of phenotypes will approach the normalized vector ${\bf q}_0$ in the null-space of $H$ (the Perron-Frobenius theorem ensures this is unique), while the top eigenvalue is given by the average of of the growth rates with respect to ${\bf q}_0$, which is
\begin{equation}
\Lambda = \frac{h_{1,2}}{h_{2,1}+h_{1,2}} \Lambda_1 +  \frac{h_{2,1}}{h_{2,1}+h_{1,2}} \Lambda_2. 
\end{equation}
Since the elements of $q_0$ are less than or equal to $1$, the population growth rate is greater in the former case where the switching rates are relatively small (assuming their ratios are fixed). In other words, fitness is optimized when all the cells are concentrated in the faster growing state, as we would expect. It is only in a varying environment or when considering transient dynamics that a population reaps the benefits of phenotypic diversity.

 Kussell et al. analyzed the situation where the environment is switching \cite{kussell2005}, with the goal of understanding how the phenotypic switching rates which optimize the populations long-term growth rate are related to the environmental switching rates.  We illustrate their results within the context of the two state example above and assume that the environment can stochastically switch between two states at a rate $w$, with states $1$ and $2$ being optimal for phenotypic states $1$ and $2$, respectively; see Figure \ref{fig:3} (A). Additionally, we make the assumption that the phenotypic switching rates are symmetric ($h = h_{1,2} = h_{2,1}$). In this model, there are now two matrices $G$, $G(1)$ and $G(2)$, whose diagonal entries are the growth rates in the corresponding environmental state.  In order to analyze the long-term dynamics, Kussell et al. consider how the state of the population, ${\bf N} =(N_1,N_2)^T$, evolves between the times $t_j$ when the environment changes state. The state of the system at a $t>t_{j-1}$ can be obtained by solving Equation \eqref{Nk_switching} with the initial condition ${\bf N}(t_{j-1})$. 
After the population has spent a long duration in environment $k_j$, the dynamics are as described above: The distribution of phenotypes approaches the eigenvector of $A(k_j) = G(k_j) + H$, ${\bf q}_0(k_{j})$, corresponding the top eigenvalue, $\lambda_0(k_j)$. If the environmental transitions are sufficiently far apart, the state of the population at the beginning of the $j$th environmental epoch is either $N_{\rm tot}(t_{j}){\bf q}_0(1)$ or $N_{\rm tot}(t_{j}){\bf q}_0(2)$, which can be taken as the initial conditions for Equation \eqref{Nk_switching} in order to obtain the number of cells at the end of the next epoch,  $N_{\rm tot}(t_{j+1})$. After some linear algebra, we obtain an iterative equation for the total number of cells at the beginning of each epoch. In the special case where the phenotypic switching rates a symmetric, this has the simple form
\begin{equation}\label{Nj}
N_{\rm tot}(t_{j+1}) \approx N_{\rm tot}(t_{j})  e^{\lambda_0(k_j)T_j}{\bf q}_0(1)^T{\bf q}_0(2).
\end{equation}
Here, $T_j = t_{j+1} -t_{j}$ is the time between the $j+1$ and $j$th environmental transitions, which is assumed to be large relative to the switching rates.
 The vector product on the right hand side arises from projection of previous epoch's steady-state phenotype distribution onto the new epoch's steady-state phenotype distribution. Note that the entries of ${\bf q}_0(k_{j})$ are the proportions of the phenotypes which optimize growth in environment $k_j$; when these proportions are reached the population grows as $\lambda_0(k_j)$. Thus ${\bf q}_0(1)^T{\bf q}_0(2)$ tells us how ``far" the two epoch's optimal phenotype distributions are from each other.

Notice that Equation \eqref{Nj} has a similar structure to Equation \eqref{Nk_cohen} from the previous section, with the factor $f_j = e^{\lambda_0(k_j)T_j} {\bf q}_0(1)^T{\bf q}_0(2)$ playing the role of the arithmetic fitness.  However, unlike the equations of the previous section, in Equation \eqref{Nj} the multipliers $f_j$ are not drawn independently, instead correlations emerge due to the dependence on the previous environmental epoch's phenotype distribution. 
 Nonetheless, we can still iterate Equation \eqref{Nj} to obtain
\begin{equation}\label{Nj_iter}
N_{\rm tot}(t_j) \sim e^{\sum_j \left[ \lambda_0(k_j)T_j  + \ln  {\bf q}_0(1)^T{\bf q}_0(2) \right]  } = e^{T \Lambda},
\end{equation}
which (after some simplification), reveals the long-term fitness $\Lambda$ to be
\begin{align}\label{switching_Lambda}
\begin{split}
\Lambda &=   \lambda_0(1) \rho_1 +  \lambda_0(2) \rho_2+ \frac{1}{\tau} \ln  {\bf q}_0(1)^T{\bf q}_0(2).
\end{split}
\end{align}
Here, $\rho_k$ is the steady-state distribution of environmental states (or said differently, the fraction of time spent in environment $k$) and $\tau$ is the average duration of an environmental epoch, or $\langle  T_j\rangle$. 
The term  $\lambda_0(1) \rho_1 +  \lambda_0(2) \rho_2$ is the average over the population growth rates that would be reached if the population were to spend an infinite amount of time in each environment; however, since the environmental epochs are finite, the population's rate of growth will never converge to the long-term growth rate in a given environment. The final term captures the penalty that the population pays for not having its cells concentrated in the optimal state when they enter a new environment. This term will be a large negative number when ${\bf q}_0(1)$ and ${\bf q}_0(2)$ are pointing in nearly orthogonal directions (i.e., when phenotypes which are highly beneficial in environment $k$ grow slowly in environment $k'$). In this case, when the environment switches the population will grow very slowly at first before eventually converging to the optimal growth rate $\Lambda({k'})$. To make $\ln {\bf q}_0(1)^T{\bf q}_0(2)$ small, the population can switch rapidly between phenotypes, since the eigenvectors ${\bf q}_0(k)$ will depend only weakly on the state $k$ in this limit. However, this rapid switching will simultaneously have the effect of reducing the first term in Equation \eqref{switching_Lambda}. Notice that the final term in Equation \eqref{switching_Lambda} is weighted by the average duration of the environment, so that the penalty for not diversifying the population becomes negligible when the environmental transitions are extremely rare. 

From the calculation shown above, we learn that in order for populations to optimize their fitness they must balance a tradeoff between growing as fast as possible in each environment, and maintaining the ability to adapt to new environments. While switching between different phenotypes tends to reduce the fitness in each environment, it allows populations to adapt to new environments more quickly. But at what point do the benefits of switching outweigh the deleterious effects? Kussell et al. optimized Equation \eqref{switching_Lambda} to show that the switching rates that result in the highest long-term growth rate are synchronized to the environmental switching rates, i.e. growth is maximized when $h = w$. More generally, they show that if the environment switches at a rate $w_{i,k}$ between states $i$ and $k$, the population will grow fastest by switching between phenotypes $i$ and $k$ at rates $h_{i,k} = w_{i,k}$. This is similar to Kelly's gambling example, where profits are maximized by gambling a fraction that is linearly related to the the probability of winning a single round. 


\section{Phenotypic variability in constant environments}

\subsection{Age-structured models}\label{sec:age_struct}
\begin{figure}[h!]
\centering
\includegraphics[scale=0.9]{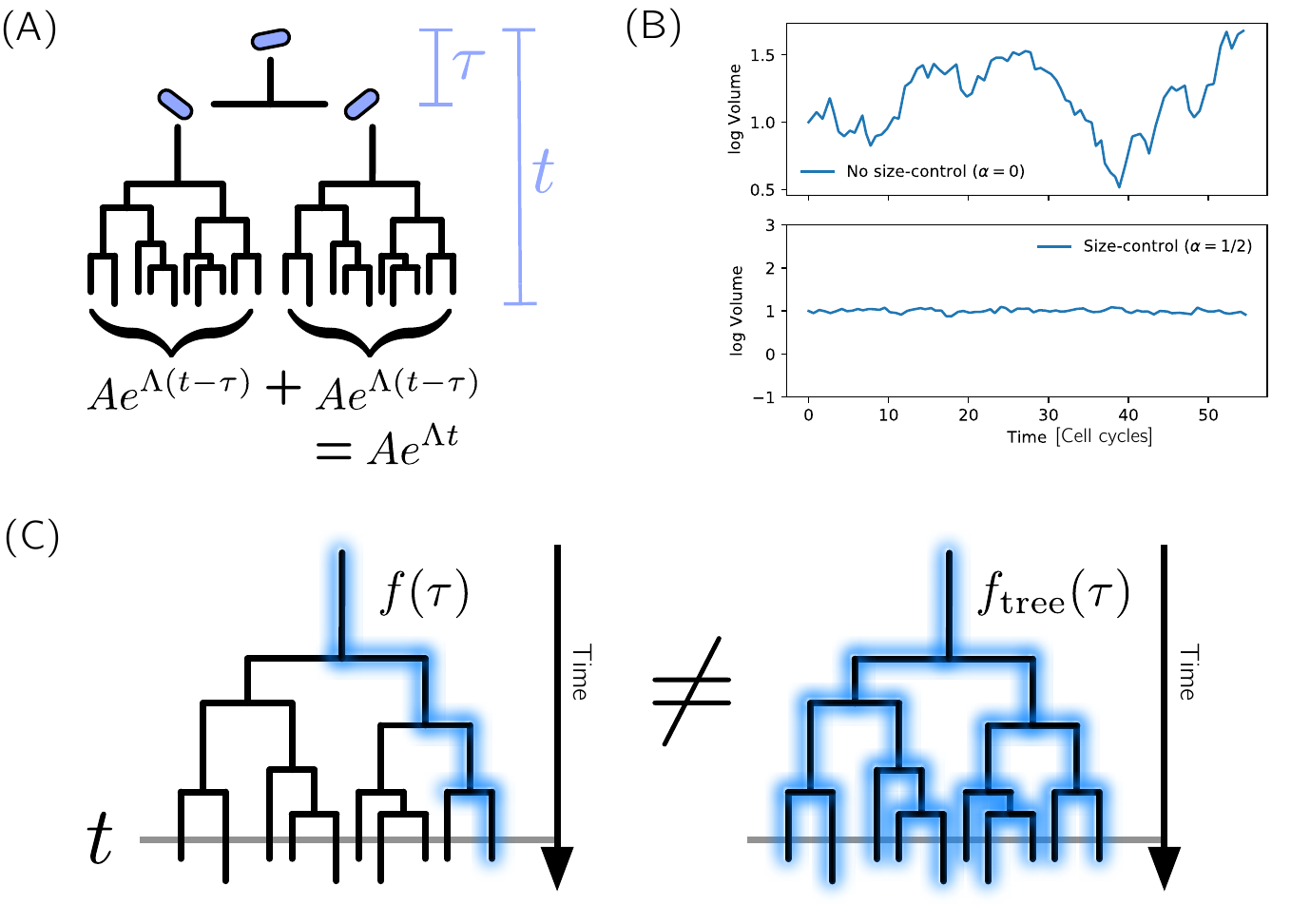}
\caption{
(A) An illustration of the recursive approach used to obtain the population growth rate within the context of the Powell's model. Because the two sub-trees spawning from the initial two daughter cells are self-similar, their sizes must be identical on average.   (B) The effects of cell-size regulation on the fluctuations in initial cell volume along a lineage. In the top panel a simulation is performed with the cell-size control parameter $\alpha$ set to zero. Here, we see large fluctuations of the cell volume at the beginning of the cell cycle. These fluctuations will be a byproduct of adding exponential volume accumulation at the single-cell level to the model proposed by Powell. In the second panel, simulations are performed with $\alpha=1/2$ and we see a stable distribution of cell-sizes. Time is in units of cell cycles. (C) The distinction between sampling generation times along a lineage $f(\tau)$ and throughout the entire population tree, $f_{\rm tree}(\tau)$. To obtain $f(\tau)$, we simple follow any lineage by selecting a single daughter cell at each branch in the tree and make a histogram of all generation times obtained along this lineage. To obtain $f_{\rm tree}(\tau)$, we record the generation time of every cell in the entire tree, including the generation times of those cells that are present when we observe the population.   }\label{fig:4}
\end{figure}

 In the previous sections we saw how organisms can hedge their bets against changing environments by diversifying the population. Increased fitness in the long-term came at the expense of fitness over a single epoch. However, some variation in any population is inevitable, and it is therefore interesting to understand how different sources of variability affect fitness in a \emph{constant} environment. We begin with a biologically incorrect, yet simple example of how cell-to-cell variability might manifest in a constant environment. Consider a population of cells where each cell lives for a time $\tau$, called the generation time, before dividing into two new cells; see Figure \ref{fig:4} (A). For most single-celled organisms, $\tau$ has a narrow distribution around some mean $\tau_0$. If the distribution is so narrow that any deviation from $\tau_0$ can be neglected, then the number of cells at a time $t$ will be $N(t) = N(0)2^{t/\tau_0}$ since $t/\tau_0$ is the number of generations. It follows that the long-term growth rate is $\Lambda = \ln(2)/\tau_0$. How might the inevitable variation around $\tau_0$ affect $\Lambda$? 
 
 This question was addressed by Powell  \cite{powell1956}, although the underlying mathematics had been developed earlier by Euler and Lotka  \cite{Baca2011}.  Powell assumed that the generation time of each cell is drawn independently from a distribution $f(\tau)$ when the cell is born. Empirically, $f(\tau)$ can be obtained as the histogram of generation times from all cells along a single-lineage, or all the cells in the population.  In this case, the total number of cells $N$ will grow exponentially at some rate $\Lambda$. Starting from one cell, we have $\langle N(t) \rangle = Ae^{\Lambda t}$ where $A$ is a constant which accounts for the transient dynamics of the population and the average is taken over many realization of the population. Only in the special case where division times are exponentially distributed will $A$ be equal to $1$. To relate $f(\tau)$ to $\Lambda$, we will follow the recursive derivation of ref. \cite{lin2017}, which begins by supposing that the first cell has a generation time of $\tau$. Since the daughter cells' generation times are independent and identically distributed (to each other and the first cell), conditioned on $\tau$ they will each produce a progeny population that has an average size $Ae^{\Lambda (t-\tau)}$ at time $t$.  It follows that the average size of each daughter's progeny will be
 \begin{equation}
A\int_0^{\infty} e^{\Lambda (t-\tau)} f(\tau)d \tau.
 \end{equation}
Expressing $\langle N(t) \rangle$ in terms of the average size of these progenies yields 
 \begin{equation}
 \langle N(t) \rangle = Ae^{\Lambda t}  = 2A\int e^{\Lambda(t-\tau)}f(\tau)d\tau.
 \end{equation} 
 Dividing both sizes by $2A e^{\Lambda t}$ yields the so-called \emph{Euler-Lotka} equation \cite{powell1956,Jafarpour2018,lin2020,levien2020}: 
 \begin{equation}\label{EL}
 \frac{1}{2} = \int_0^{\infty}f(\tau)e^{-\Lambda \tau}d\tau.
 \end{equation}
 The argument is illustrated graphically in Figure \ref{fig:4} (A).
 Note that when $f(\tau) = \delta (\tau-\tau_0)$ we retrieve the result for deterministic generation times, $\Lambda = \ln(2)/\tau_0$. In the other limit, when cells divide at a constant rate $\lambda$ which is independent of the cell's age, $f(\tau) = \lambda e^{-\lambda \tau}$ and we obtain $\lambda = \Lambda$. In this case, the mean generation time is $\langle \tau \rangle  = 1/\lambda$. Comparing the two population growth rates, we see that the growth rate is reduced by a factor of $\ln(2)$ when generation times are deterministic, suggesting variability has a beneficial effect on fitness. Indeed, for a fixed $\tau_0$, increasing the variation around $\tau_0$ will result in a larger population growth rate compared to the deterministic case.  Given a narrow Gaussian distribution of generation times with mean $\langle  \tau\rangle$ and variance $\sigma_{\tau}^2$, we can solve Equation \eqref{EL} explicitly to obtain \cite{lin2018,levien2020b}
    \begin{equation}\label{EL_Lambda}
  \Lambda = \frac{2 \ln(2) /\langle \tau \rangle }{1 + \sqrt{1 - 2 \ln(2) \sigma_{\tau}^2/\langle \tau \rangle^2}},
  \end{equation}
  which is an increasing function of $\sigma_{\tau}^2$,  as expected.

The increased growth is a result of an asymmetry between the beneficial effects of cells with generation times that are shorter than the average, and the deleterious effects of cells with longer than average generation times. To see this, suppose the  population originates from a single ancestral cell with a generation time $\tau \ne \langle \tau \rangle$. At some later time, the population size will differ from a population in which the first cell had a generation time $ \tau = \langle  \tau\rangle$ by a factor $e^{\Lambda (\tau-\langle \tau \rangle)}$. Due to the convexity of the exponential function, this factor is not symmetric around $\langle \tau \rangle$ and averaging over all possible initial generation times yields a factor greater than $1$: 
 \begin{equation}
  \left \langle e^{\Lambda (\tau-\langle \tau \rangle)}\right \rangle> e^{\Lambda (\langle \tau \rangle-\langle \tau \rangle)}  = 1.
  \end{equation}
  (This is an example of Jenson's inequality \cite{gardiner2009}.) It follows that the deviations of the initial cell's generation time from $\langle \tau \rangle$ will, on average, make the population size larger. Since the same can be said for any of the cells (recall that they are all independent and identically distributed), it is clear why variation in generation times around a fixed mean is favorable.   

The preceding argument might lead us to conclude that evolution will select for cells with a broad distribution of generation times (at least in constant environments), yet in \emph{E. coli} the CV of generation times is only around $0.29$ \cite{lin2017}. However,  the conclusion that variability is beneficial stems from an overly simplistic view of single-cell dynamics, which neglects two well established experimental observations: The distribution of cell sizes is heavily regulated and single cells accumulate biomass exponentially  \cite{lin2017}. These two observations are incompatible with a model where generation times are uncorrelated between mother and daughter cells, because small fluctuations in generation times will result in unbounded growth of the size distribution. 
To see this, consider the  initial volume $v_0$ of a daughter cell with a mother cell that has volume \footnote{Here, volume is a proxy for size. For a rod shaped organism like \emph{E. coli}, we could have just as easily used length or mass} $v_0'$. Assuming cells divide symmetrically, 
 \begin{equation}
v_0 = v_0'e^{\lambda \tau}/2
 \end{equation}
 where $\lambda$ is the (specific) growth rate (defined as $(1/v)dv/dt$) and $\tau$ the generation time of the mother cell. Taking logarithms on both sides, we obtain  \cite{amir2014}
 \begin{equation}\label{csr_lnv0}
 \ln v_0 =-\ln(2) +  \ln v_0' + \lambda \tau.
 \end{equation}
 Since cells double in size on average, it must be that $\langle e^{\lambda \tau } \rangle =2$, implying $\langle \lambda \tau \rangle \approx \ln 2$. Therefore, Equation \eqref{csr_lnv0} becomes
 \begin{equation}
  \ln v_0 \approx \ln v_0' + \xi.
 \end{equation}
 The noise term $\xi$ is a result of the combined effects of variation in growth rate and generation time. 
 If the generation time and growth rate are both independent of the size, $\ln v_0$ undergoes a random walk so that  the population fails to reach a stable size distribution; see Figure \ref{fig:4} (B). Coupling between the final size and size at division is therefore necessary in order to keep the size distribution from growing indefinitely along a lineage.  Since the size distribution is typically small in microbes, it is common to use a linear model for the coupling between final and initial size, which is parameterized by a cell-size control parameter $\alpha$ and a size increment $\Delta$. This takes the form
 \begin{equation}\label{csr-model}
 v_f = 2 \alpha v_0 + 2(1-\alpha) \Delta + \xi.
 \end{equation} 
The parameter $\alpha$ determines the cell-size regulation strategy: When $\alpha=1$  (known as a ``sizer" strategy), cells divide at a critical size, while when $\alpha =1/2$ (known as an ``adder" strategy) cells add a constant size $v_0$ between birth and division.  We refer to refs \cite{ho2018} for an in-depth discussion of the cell-size control model and its justification.

Using the cell-size regulation model,  Lin and Amir explore whether variability in various quantities increases or decreases the population growth rate. While it is difficult to obtain the growth rate analytically, an intuition can be gained by looking at the limit where there is no variation in growth rates. If each cell in the population grows at exactly $\lambda_0$, then between cell divisions the change in the total volume of the population of $N(t)$ cells is given by 
 \begin{equation}
\frac{d}{dt} V(t) = \sum_{i=1}^{N(t)}\frac{d}{dt}v_i = \lambda_0 V(t).
 \end{equation}
 That is, the growth rate of the volume of the colony is $\lambda_0$. 
 If the total number of cells grows at a rate $\Lambda$, then we can obtain the average cell size in the population at a time $t$:
 \begin{equation}
\langle v \rangle  = \frac{V(t)}{N(t)}  \propto e^{(\lambda_0 - \Lambda)t},
 \end{equation}
 but since cell size is controlled, $\langle v \rangle$ must converge to a constant, implying $\lambda_0 = \Lambda$. Interestingly, this is independent of the volume additive noise appearing in Equation \eqref{csr-model} and the cell-size regulation parameter! In contrast, consider how we might use Equation \eqref{EL} to obtain the population growth rate when the single-cell dynamics are described by the cell-size control model. In this case, the distribution of generation times can be approximated by a Gaussian around the average generation time $\langle \tau \rangle \approx \ln(2)/\lambda_0$. Because the variation in $\tau$ depends on the cell-size regulation parameter and volume additive noise, the expression for $\Lambda$ obtained by plugging this distribution into Equation \eqref{EL} will depend on both the cell-size regulation parameter and the volume additive noise term, in direct contrast with the argument above \cite{lin2017}. 
 
 A key assumption in the derivation of Equation \eqref{EL} that is violated by the cell-size control model is that generation times are uncorrelated across generations.  In light of this observation, it is natural to ask: is there a general relation between the distribution of generation times and the population growth rate that holds in the presence of correlations between mother and daughter cells. In fact, a relationship identical to Equation \eqref{EL} holds in the presence of such correlations, with the caveat that the interpretation of $f(\tau)$ is slightly different. Specifically, $f(\tau)$ should be replaced with $f_{\rm tree}(\tau)$, the distribution of generation times throughout the entire population tree. When generation times are uncorrelated, it can be shown that $f_{\rm tree}(\tau) = f(\tau)$; see Figure \ref{fig:4} (B).

 \subsection{Variability in growth rates}\label{sec:growthratevar}
 Using the generalization of the Euler-Lotka equation, Lin and Amir have analyzed the cell-size regulation model with noise in growth rates, generation times and volumes \cite{lin2020}. In order to add noise to the single-cell growth rates, they suppose that each cell in the population grows at a constant rate throughout the cell-cycle which has the form 
 \begin{equation}\label{ar1lambda}
 \lambda = \lambda'\rho + (1-\rho)\langle \lambda \rangle + \xi_{\lambda}
 \end{equation}
 where $\lambda'$ is the mother cell's growth rate, $\rho$ is the correlation coefficient between mother and daughter cells and $\xi_{\lambda}$ is a Gaussian noise term with mean zero. Given the argument made above for the case when there is no variation in the growth rates, it should not be surprising that to leading order (in the noise parameters) the population growth rate depends only on the variation and mother-daughter correlations in growth rates, and not the cell-size regulation parameter or other sources of noise.  In particular, the correlations increases the population growth rate while variation in growth rates decreases the population growth rate. This is shown in ref. \cite{lin2020} using a generalization of the recursive approach discussed above.
 
  The negative effects of growth rate variation within the context of Equation \eqref{ar1lambda} can be understood by considering the case where $\rho=0$ and supposing cell size is perfectly regulated (so that all cells divide at exactly twice their birth size). In this limit, generation times are uncorrelated so that we can apply Powell's theory.   To do so, we map the variation in growth rates to variation in generation times and apply Equation \eqref{EL_Lambda}. Starting with the average generation time, we have
 \begin{equation}
\langle  \tau \rangle  =\left \langle \frac{\ln(2)}{\lambda} \right \rangle =\left \langle \frac{\ln(2)}{\langle \lambda \rangle + \xi_{\lambda}}  \right \rangle  \approx \frac{\ln(2)}{\langle \lambda \rangle}\left( 1+ \frac{\sigma_{\lambda}^2}{\langle \lambda \rangle^2} \right),
 \end{equation}
 where $\sigma_{\lambda}^2$ is the variance in single-cell growth rates taken over all cells along a lineage. 
The variation in growth rates therefore increases the average generation time by a factor of $1+{\rm CV}_{\lambda}^2$. 
While the variation in growth rates will also change the variance in generation times, the effect on the average generation time will have a more pronounced effect on the population growth rate $\Lambda$. Indeed, plugging the expressions for $\langle \tau \rangle$ and $\sigma_{\tau}^2$ (which we do not derive explicitly here) into Equation \eqref{EL_Lambda} and Taylor expanding in $\sigma_{\lambda}^2$ yields \cite{lin2020,lin2017,thomas2017}
 \begin{equation}\label{Lambda_constant_gr}
 \Lambda = \langle \lambda  \rangle\left( 1 - \left[1-\frac{\ln(2)}{2} \right]\frac{\sigma_{\lambda}^2}{\langle \lambda \rangle} \right).
 \end{equation}
This example  illustrates the importance of specifying what phenotypes the variability is taken to be in: In a model where each cell is independently assigned a random growth rate and cells divide upon reaching a critical size, variation around a fixed average growth rate is deleterious. However, in this model generation times are also independent between cells.  Therefore, Powell's results tells us that if it is not the average growth rate that is fixed, but rather the average generation time, variability will be beneficial!

\begin{figure}[h!]
\centering
\includegraphics[scale=0.9]{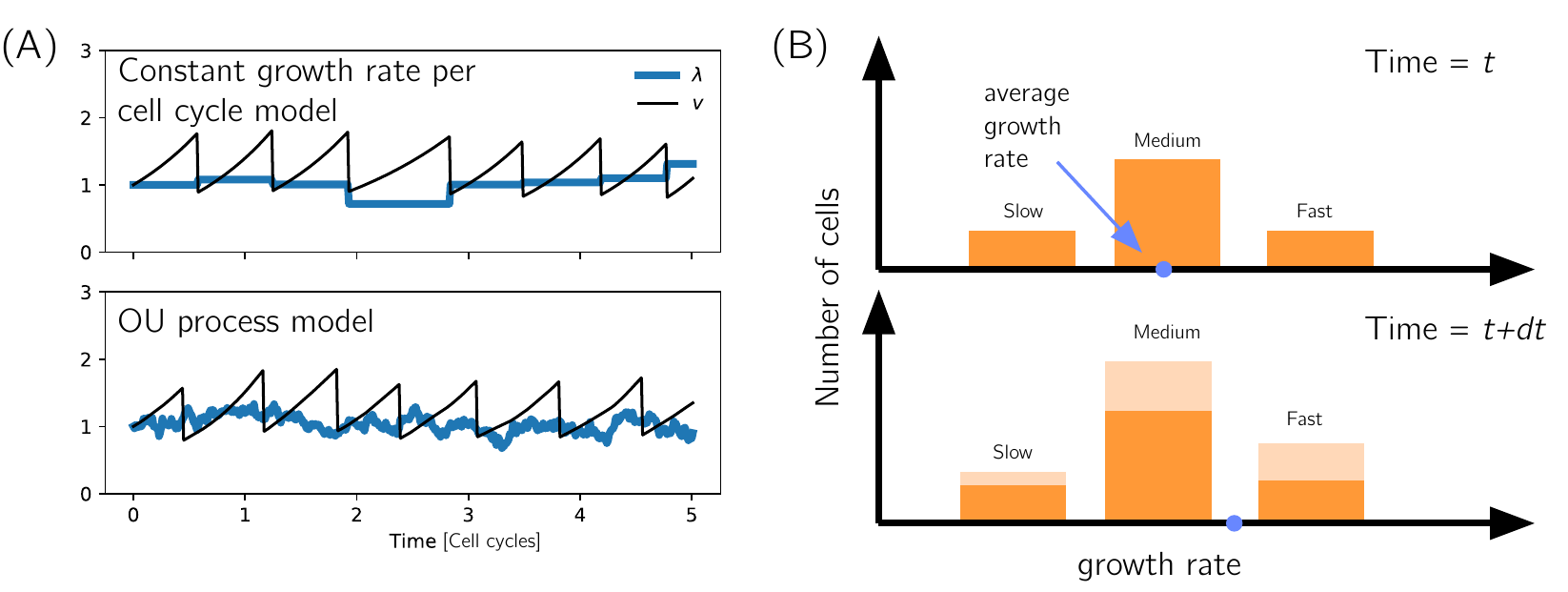}
\caption{
(A) Two models of growth rate variation. In the model used by Lin and Amir (top) each cell has a different growth rate which is constant over the cell cycle. In the OU process model (bottom), the growth rate changes continuously and cell divisions are invisible from the perspective of the growth rate fluctuations. Time is in units of cell cycles. (B) An illustration of the intuition underlying the beneficial effects of growth rate variation in the OU process model and Fisher's fundamental theorem. The top panel shows a histogram of a population at time $t$, with the growth rates binned into slow, medium and fast groups. The histogram is symmetric so that the average growth rate is in the center of the medium group. If growth rates are perfectly heritable, then at time $t+dt$ the average will have increased, since the fast growing cells produce more offspring.   }\label{fig:5}
\end{figure}

In Lin and Amir's analysis, it is assumed that growth rates are constant throughout the cell-cycle, so variation in growth rates is introduced between cell cycles. A plausible alternative model assumes growth rates undergo an Ornstein-Ulhenbeck process; that is 
 \begin{equation}\label{growth_OU}
 \frac{d}{dt}\lambda = \gamma(\lambda_0-\lambda) + \xi_{\lambda}.
 \end{equation}
Here, $\lambda_0$ is the average growth rate along a lineage, $1/\gamma$ is the relaxation time-scale and $\xi_{\lambda}$ is a white noise term with variance $2D_{\lambda}$. 
 Equation \eqref{growth_OU} captures basic qualitative aspects of the growth rate dynamics, namely, that growth rates fluctuate around a mean and are approximately Gaussian. In the OU process model, the division events are invisible from the perspective of the growth rate fluctuations; that is, when a cell divides its daughter simply inherits the mother cell's instantaneous growth rate which continues to evolve according to Equation \eqref{growth_OU};  see Figure \ref{fig:5} (A).

The OU model for growth rate variation was previously analyzed in refs \cite{Nicola2008,mora2013}, where the dynamics are motivated by fluctuations in growth limiting protein concentrations. Here, we roughly follow the derivation in ref \cite{mora2013}.  In order to obtain the long term growth rate in terms of the model parameters, we once again notice that because cell-size is regulated, the average growth rate in a snapshot of the population, which we will denote  $\langle \lambda \rangle_p$,  is equal to the long-term exponential growth rate of the population, $\Lambda$. (Note that $\langle \lambda \rangle_p$ is in general distinct from the average growth rate along a lineage, $\lambda_0$.) In order to derive an equation for $\langle \lambda \rangle_p$, we define $N(t,\lambda)$ as the number of cells with growth rates in $[\lambda,\lambda+d\lambda)$ at time $t$. For simplicity, we assume that cell-size is strongly regulated so that any variations in the cell-size at division can be neglected. In this case, the dynamics of $N(t,v,\lambda)$ can be described by the transport equation
\begin{equation}\label{fpN}
\frac{\partial}{\partial t}N(t,\lambda) =\lambda N(t,\lambda)  + \frac{\partial}{\partial \lambda}[\gamma(\lambda_0-\lambda)N(t,\lambda)] + D_{\lambda}\frac{\partial^2}{\partial \lambda^2}N(t,\lambda).
\end{equation}
Here, the first term on the right-hand side represents the increase in the number of cells due to cell division. To get some intuition for this term, observe that if all cells have the same growth rate, $\lambda$, and cell-size is regulated, the population growth rate will also be $\lambda$.  The second term represents the deterministic dynamics of $\lambda$, which relaxes to the mean $\lambda_0$ at a rate $\gamma$. Finally, the last term term captures the diffusion in $\lambda$ due to stochastic fluctuations \cite{gardiner2009}. We can express $\langle \lambda \rangle_p$ in terms of an integral over $N(t,\lambda)$ as
\begin{equation}
\langle \lambda \rangle_p  =  \Lambda = \frac{1}{N_{\rm tot}} \int_0^{\infty}\lambda N(t,\lambda)d\lambda. 
\end{equation}
In the long-term, $\langle \lambda \rangle_p$ will converge to a constant, the value of which can be determined by studying the transient dynamics.  Differentiating $\langle \lambda \rangle_p$ and performing some straightforward simplifications leads to 
\begin{align}\label{dlambdapdt}
\begin{split}
\frac{d}{dt}\langle \lambda  \rangle_p &= \gamma (\lambda_0 - \langle \lambda  \rangle_p) + \underbrace{ \langle \lambda^2 \rangle_p -  \langle \lambda  \rangle_p^2}_{= \sigma_{\lambda,p}^2}.\\
\end{split}
\end{align}
By setting Equation \eqref{dlambdapdt} to zero and solving for $\langle \lambda  \rangle_p$, we obtain $\Lambda$:
 \begin{equation}\label{Lambda_ou}
 \Lambda = \lim_{t \to \infty}\langle \lambda  \rangle_p =\lambda_0+  \sigma_{\lambda,p}^2/\gamma.
  \end{equation} 
Thus, variability is always beneficial, regardless of the relaxation time-scale $\gamma^{-1}$. This result may seem contradictory to Equation \eqref{Lambda_constant_gr}, which is derived under the assumption that growth rates are uncorrelated across generations and constant over the cell cycle.  In particular, why don't we retrieve the behavior of the model in which growth rate correlations are absent by taking $\gamma \to \infty$? 
  The resolution to this apparent paradox lies in the observation that, in the OU process model, $\sigma_{\lambda,p}$ is the variability in \emph{instantaneous} growth rates in a snapshot of the population. This is in fact a fundamentally different quantify than $\sigma_{\lambda}$ in Equation \eqref{Lambda_constant_gr}, which is the variation in growth rates taken over all cells along a lineage.


To gain an intuition for why variation in the instantaneous growth rate increases the population growth rate, it is useful to consider the limit where $\gamma$ is very small. In this limit, the time-scale over which growth rates fluctuate is much larger than the generation time of a cell. Imagine starting with a population of cells at time $t$ with some distribution of growth rates defined by a variance $\sigma_{\lambda,p}^2$ and mean $\langle \lambda \rangle_p$. Now consider the average growth rate at some later time  $t+dt$. How will the average growth rate at this later time be related to $\langle \lambda \rangle_p$? Each cell will grow and divide to produce offspring which inherit their growth rate, but the cells with large growth rates will produce more offspring. Thus, the distribution at $t+dt$ will be biased towards the faster growing cells, implying $\langle \lambda \rangle_p(t+dt) > \langle \lambda \rangle_p(t)$. This intuition is illustrated in  see Figure \ref{fig:5} (B).  If we increase $\sigma_{\lambda,p}^2$, then there will be more cells with both larger and smaller growth rates; however, since the progeny's grow exponentially, the cells with larger growth rates will contribute disproportionately the the average. For this reason, increasing $\sigma_{\lambda,p}^2$ will lead to a larger increase in average growth rate.  We can now see why variation increases population growth for any value of $\gamma$: The larger the variation, the more likely it will be that the population produces cells with very high growth rates.  These cells contribute disproportionately to the population growth rate. Notice the similarity between the intuition used here and that applied to the independent generation time model in the previous section. In both cases, the asymmetry between the contribution of fast and slow growing cells in an exponentially proliferating population plays a key role. This effect is related to Fisher's fundamental theorem, which states that in a population where fitnesses (or the exponential growth rates) are perfectly heritable (i.e., $\gamma \to 0$), the increase in fitness over a generation is proportional to the variation in fitness \cite{fisher1931}. 

Why doesn't the intuition described above apply to the model analyzed by Lin and Amir in which growth rates are constant over the cell cycle? The answer is that when growth rates are fixed over the cell cycle, faster growing cells are more likely to divide (recall that the regulation of cell sizes guarantees faster growing cells must have shorter generation times). If growth rates are uncorrelated, these faster growing cells will produce a progeny with growth rates that are not, on average, any faster than the typical cell in the population. On the other hand, slow growing cells will be less likely to divide in any given time interval, since they must grow for longer to reach the target size. 
The result is that the entire population's progeny will be biased towards slower growing cells, rather than faster growing cells as is the case in the OU process model.

  \subsection{Aging and asymmetric segregation \label{3.3}}
  Thus far we have considered variability that is introduced by stochastic fluctuations in single-cell traits, such as growth rates and generation times, but variability may also be introduced by deterministic dynamics. One such example is  \emph{biological aging}, defined as the deterioration of organismal function with time. It was once thought aging was a phenomenon restricted to multicellular organisms and that microbes are immortal, in the sense that any lineage of cells undergoing binary fission can proliferate indefinitely. However, detailed experimental studies of single-cell organism have revealed that many cells divide asymmetrically, such that one of the progeny cells always receives the older, damaged biomolecules. Eventually, the lineages which accumulate the older molecules will decline in fitness.  Other symmetrically dividing cells also experience senescence (process of deterioration with age) due to damage in macromolecules such as DNA or proteins \cite{coelho2013fission, coelho2014fusion, lindner2008asymmetric, stewart2005, Vaubourgeix2015, Vedel2016}. Recently, it was also observed that some beneficial cellular components can segregate in different ratios and lead to heterogeneous fitness distribution \cite{bergmiller2017biased, savage2010spatially}. In a recent theoretical paper, Lin et al. tried to quantify the effects of such asymmetry between the daughter cells in the limit where the ratio between cellular components is deterministic \cite{lin2019optimal}. In this section, we introduce their model and main results. In the next section, we show how the self-consistent equation explored by Levien et al. \cite{levien2020} can be used to generalize the work of Lin et al. \cite{lin2019optimal}.

Lin et al. assume that the amount of key proteins, $D$, increases proportionally with the volume:
\begin{align}
D(V) = D_b + S(V - V_b),
\end{align}
where $S$ is a ``stress'' parameter that sets the accumulation rate, while $D_b$ and $V_b$ are the protein level and volume at cell birth, respectively.  The growth rate is assumed to be a Hill-function of the concentration of the key protein ($\sigma = D / V$); that is, 
\begin{align}
	\lambda[\sigma] = \frac{\lambda_0 + \lambda_1 \sigma^n}{1+\sigma^n}.
\end{align}
Here, $\lambda_0$ is the growth rate of a cell with no key protein and the sign of $\lambda_1$ determines whether the growth rate increases or decreases as the protein accumulates. When the cell divides, $\frac{1+a}{2}$ of the key proteins go to one daughter cell and $\frac{1-a}{2}$ goes to the other cell: $a = 0$ corresponds to a completely symmetric segregation and $a = 1$ to a perfectly asymmetric one. One important finding from the paper is that whether the population growth rate $\Lambda_p$ is an increasing or decreasing function of asymmetry $a$ depends on the concavity of $\lambda[\sigma]$:
\begin{align}
	\Lambda_p = \lambda[S] \left(1 + \frac{1}{4\ln 2} \frac{\lambda''[S]}{\lambda[S]}S^2 a^2\right).
\end{align}
Another interesting point is that the deleterious and beneficial key proteins exhibit qualitatively different behaviors. Specifically, the optimal asymmetry $a_c$ that maximizes $\Lambda_p$ undergoes smooth second-order phase transition if the key proteins are beneficial, whereas the transition is first-order if the key proteins are detrimental to growth.

  \subsection{Mathematical generalizations}
 
The examples above illustrate the subtle mathematical structure underlying population dynamics of phenotypically heterogeneous populations, showing that  the effects of variability are sensitive to the source of the variation in the relevant phenotypic traits. It is therefore natural to ask whether these results can be placed within a unified mathematical framework. To this end, Levien et al. \cite{levien2020} have considered a general modeling framework allowing the study intrinsic variability in an arbitrary phenotypic trait, or combination of traits, which is represented by the vector ${\bf x} = (x_1,\dots,x_l) \in {\mathbb R}_+^l$.  Here, it assumed that each cell is born with a value of ${\bf x}$ drawn from a distribution $f({\bf x}|{\bf x}')$, which depends on the phenotype ${\bf x}'$ of the mother cell. Note that ${\bf x}$ may not change over the cell-cycle. For example, ${\bf x}$ could represent the initial concentration of a protein, or some macroscopic phenotype, such as the average growth rate over the cell cycle. However, it could not represent the age or volume of a cell, since these change over the course of the cell cycle. Within-cell-cycle variation can be incorporated into the transition function $f({\bf x}|{\bf x}')$ or bad adding additional components to ${\bf x}$ for different points in the cell-cycle.   After a time $\tau({\bf x})$, the cell divides and produces two new cells with phenotypes drawn from $f({\bf x}|{\bf x}')$ and generation times depending on these phenotypes. This framework can, in principle, describe any possible form of \emph{intrinsic} (as opposed to environmental) \emph{generation time} variability through the appropriate choices of the phenotype ${\bf x}$ and distribution $f$. Therefore, existing models can be derived as specific cases of this framework. These include models where division is asymmetric \cite{aldridge2012,chao2016,lin2019optimal,Jafarpour2018} and where the relationship between successive generation times is non-monotonic and nonlinear \cite{mosheiff2018}.

In order to derive a generation relationship between the distribution of phenotypes and fitness, Levien et al. consider the joint distribution of ages $u$, and phenotypes ${\bf x}$ in the population, denoted $\psi(t,{\bf x},u)$. Note that $u$ refers to the chronological age of a cell (the time since birth), and does not necessarily reflect the biological age discussed in the previous section. When $u$ is less than the generation time $\tau({\bf x})$ of a cell, the dynamics of $\psi(t,{\bf x},u)$ are simple, as only two events change this distribution. First, cells will age at a constant rate. Second, as new cells are born the distribution will be \emph{diluted} by the newborn cells, causing the fraction of cells with phenotype ${\bf x}$ and age $u$ to decrease. The rate of decrease will be equal to the rate at which new cells are born, which is exactly the population growth rate, $\Lambda$. Together, these observations imply $\psi(t,{\bf x},u)$ obeys the transport equation:
 \begin{align}\label{psiFpde}
\left(\frac{\partial}{\partial t} +\frac{\partial}{\partial u}  \right)\psi(t,{\bf x},u) &= -\Lambda \psi(t,{\bf x},u).
\end{align}
This equation is known as the Von Foerster equation for an age structured population \cite{Baca2011}. Equation \eqref{psiFpde} needs to be supplemented by the boundary conditions:  
\begin{equation}\label{mvf-boundary}
\psi(t,{\bf x},0)  = 2 \int_0^{\infty}\cdots \int_0^{\infty}f({\bf x}|{\bf x}')\psi(t,{\bf x}',\tau({\bf x}'))dx_1'\cdots dx_l'. 
\end{equation}
The term $f({\bf x}|{\bf x}')\psi(t,{\bf x}',\tau({\bf x}'))dx_1'\cdots dx_l'$ represents the probability of a cell with phenotype ${\bf x}'$ dividing to produce a cell with phenotype ${\bf x}$, while the factor of $2$ comes from the fact that each cell produces two daughters.  Under balanced growth conditions, $\psi$ will converge to some time independent distribution $\psi({\bf x},u)$ describing the steady-state distribution of ages and phenotypes in a growing population. This can readily be obtained from Equation \eqref{psiFpde} by assuming a steady-state:
 \begin{align}\label{psiFpde_ss}
\frac{\partial}{\partial u} \psi(t,{\bf x},u) &= -\Lambda \psi({\bf x},u) \implies  \psi({\bf x},u) = \psi({\bf x},0)e^{-\Lambda u}, \quad u<\tau({\bf x}).
\end{align}
Here, $\psi({\bf x},0)$ is proportional to the distribution of phenotypes among all the cells that have just been born in a snapshot of the population, which we will denote $\psi_{\rm birth}({\bf x})$.  Notice that, since this distribution is time invariant, sampling those cells that have just been born at any sequence of times $t_1,t_2,\dots t_n$ yields the same distribution. Carrying out this sampling procedure at an infinite sequence of times samples every cell only once, and therefore $\psi_{\rm birth}({\bf x})$ is equivalent to the distribution of phenotypes throughout the entire history of the population including those that are currently alive. We refer to this as the \emph{tree distribution} and denote it as $\psi_{\rm tree}({\bf x})$.  Plugging Equation \eqref{psiFpde_ss} into Equation \eqref{mvf-boundary} yields a relation between the distribution of phenotypes at birth, the transition function $f$ and the population growth rate $\Lambda$ \cite{levien2020}:
 \begin{equation}\label{mvf-boundary2}
\psi_{\rm birth}({\bf x})  = 2 \int_0^{\infty}\cdots \int_0^{\infty}f({\bf x}|{\bf x}')\psi_{\rm birth}({\bf x}')e^{-\Lambda \tau({\bf x}')}dx_1'\cdots dx_l'. 
\end{equation}
Finally, integrating with respect to ${\bf x}$ and using the normalization of $f$ and $\psi_{\rm birth}$  retrieves a generalized version of the Euler-Lotka Equation \eqref{EL}: 
 \begin{equation}\label{EL2}
\frac{1}{2}  =  \int_0^{\infty}\cdots \int_0^{\infty}\psi_{\rm birth}({\bf x})e^{-\Lambda \tau({\bf x})}dx_1'\cdots dx_l'. 
\end{equation}
The exact form of the classical Euler-Lokta Equation \eqref{EL} is obtained when the phenotype ${\bf x}$ is simply the generation time, and the transition function is independent of the mother cell's generation time: $f(\tau|\tau') = f(\tau)$. We note that the special case ${\bf x} = \tau$ was studied earlier in ref. \cite{lebowitz1974}, where an exactly solvable model with correlated generation times was presented. 

Equation \eqref{mvf-boundary2} has recently been applied to calculate the population growth rate in the model of asymmetric segregation presented in Section  \ref{3.3}. In Lin et al.'s model of asymmetrically dividing key proteins, the amount of protein at birth ($\sigma_b$) is the only phenotype that determines the generation time of the cell \cite{lin2019optimal}.
\begin{equation}
    \tau[\sigma_b] = \int_1^2 \frac{1}{V \lambda[\sigma_b + S (V - 1)]}dV.
\end{equation}
Therefore, we can write down an equation of the distribution of the protein concentration at birth ($\psi[\sigma_b]$) and population growth rate $\Lambda$ as \begin{equation}\label{segregation_self_consistent}
    \psi[\sigma_b] = 2 \int_0^\infty f(\sigma_b|\sigma_b') \psi[\sigma_b']e^{-\Lambda \tau[\sigma_b']} d\sigma_b'. 
\end{equation}
Here the transition function depends on the asymmetry parameter $a$.
\begin{equation}
    f(\sigma_b|\sigma_b') = \frac{1}{2}\delta\left(\sigma_b -\frac{1+a}{2}(\sigma_b' + S) \right) + \frac{1}{2}\delta\left(\sigma_b -\frac{1-a}{2}(\sigma_b' + S) \right).
\end{equation}
In a limit of small $a$, we can reduce the problem to solving three linear equations with three unknowns ($\Lambda$, mean and variance of $\psi[\sigma_b]$), obtained from Equation \eqref{segregation_self_consistent} \cite{min2020}.

\section{Other topics and outlook}

\subsection{Finite cultures}

\begin{figure}[h!]
\centering
\includegraphics[scale=0.9]{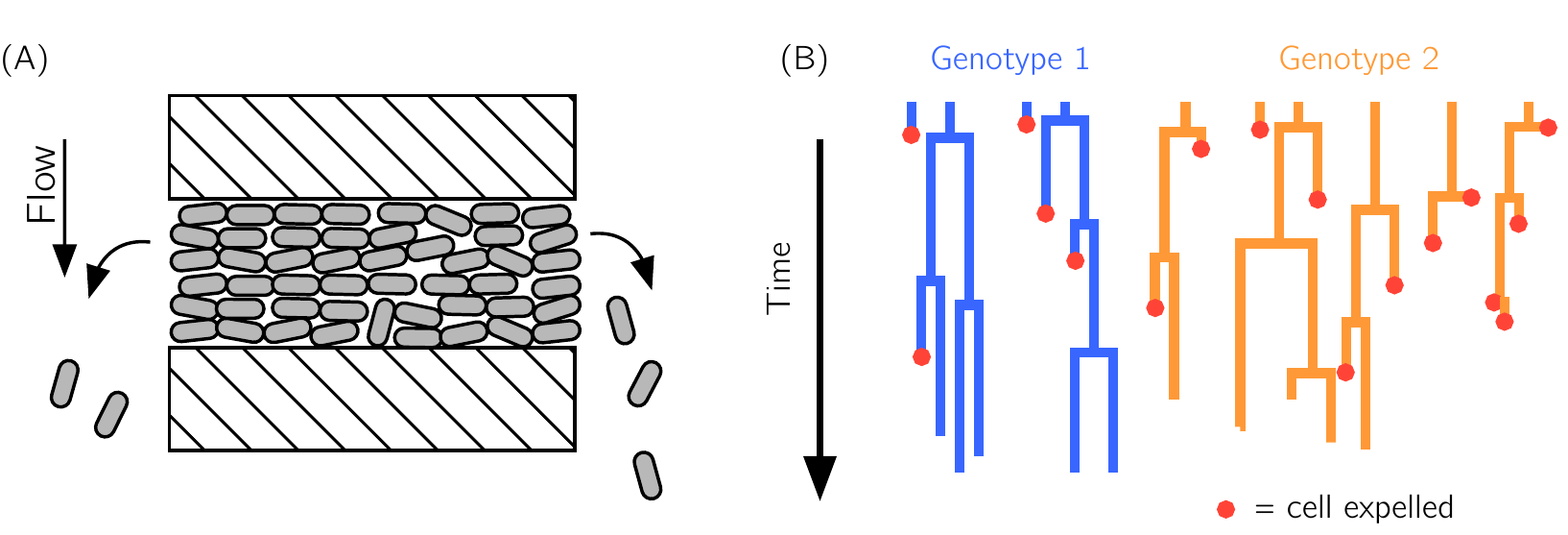}
\caption{
(A) A diagram of a microfluidic device, known as the dynamics cytometer  \cite{Hashimoto2016}, which has been used to monitor growth in finite populations of bacteria, while also tracking individual cells. (B) A diagram of the lineage tree from a Moran process with $N=10$ cells in which two genotypes compete. Each time a cell divides another cell in the population is randomly selected from all the cells in the population (regardless of which genotype has divided) and expelled. The expelled cells are indicated by red circles.  }\label{fig:6}
\end{figure}

Throughout this review it has been assumed that the populations under consideration are growing exponentially, without any bound on the number of cells. In all natural and experimental settings the assumption of uninterrupted exponential growth must be violated. To see why, simply consider that in rich growth conditions \emph{Escherichia coli} has a doubling time of roughly $20$ minutes. This means that, with each individual bacterium weighing on the order of $1$ picogram, a single day of exponential growth starting from one cell would result in a colony weighing $2^{24\times 3} \approx 4.7 \times 10^{21}$ picograms (around $5$ tons).  With finite resources available, real populations are periodically diluted so that their size remains fixed on average. For example, in experimental systems such as the microfluidic device shown in Figure \ref{fig:6} (A), or a chemostat, cells are flushed out as new media is pumped into the culture. Similarly, in natural environments such as the digestive system of a host organism, waste is periodically flushed out by the host. Perhaps the simplest mathematical model of a finite population is the \emph{Moran process}. Here, $N$ is kept fixed by selecting a cell to remove from the population at every division event; see Figure \ref{fig:6} (B). If two species are present in such a population, the principle of \emph{competitive exclusion} states that one of them will eventually dominate the population.

All of the results presented throughout this review can be generalized to the Moran process, provided the number of cells is sufficiently large. To be precise, consider two different sub-populations (which may be different species or genotypes) which, given no restriction on their sizes, will grow exponentially at rates $\Lambda_1$ and $\Lambda_2$ (the generalization to more than two species is straightforward).  Now suppose the species are grown together in a well-mixed culture that can accommodate only $N$ cells and let $N_1$ and $N_2$ be the number of these species at time $t$. We will assume that the species are not interacting, except indirectly through their competition for space in the culture. In the Moran process model, the number of cells that are expelled from the culture in a small time interval $dt$ is equal to the number of cell divisions in that interval, since each division event corresponds to a cell being expelled from the culture. It follows that $(\Lambda_1 N_1 + \Lambda_2 N_2)dt$ cells are expelled in an interval $dt$. When multiplied by $\phi_1$ and divided by $N$, this term gives us the per cell rate at which cells of species $1$ are expelled from the culture, which is
$\phi_1 [\Lambda_1\phi_1 + \Lambda_2 (1-\phi_1)]$. At the same time, new cells of type $1$ are produced at a rate $\phi_1 \Lambda_1$ per cell. Combining these terms results in the standard logistic growth equation for the species fraction $\phi_1$: 
\begin{align}\label{logistic}
\begin{split}
\frac{d}{dt}\phi_1
&=  \phi_1(1-\phi_1)(\Lambda_1 - \Lambda_2). 
\end{split}
\end{align} 
 Note that we have made no assumption about the time dependence of the rates $\Lambda_i$, nor have we made any assumption about the interspecies variability, only that each population will grow exponentially (with possibility a time dependent exponential growth rate) in the long term. Equation \ref{logistic} can be used to describe the population dynamics of two species in a constant or random environment with or without phenotypic variability within species. 

While the logistic growth, Equation \eqref{logistic}, implies that the long-term fitness of a population in a finite culture is indeed equal to the population growth rate in the exponential setting, it does not tell us how fitness is related to the distribution of phenotypes in the culture. In the previous section, we saw that $\Lambda$ is related to $f_{\rm tree}(\tau)$, the distribution of generation times over the history of the population, but since most of the cells are flushed out of the finite culture, there is no way to obtain this distribution.   What is the relevant distribution of cells in a competition experiment which determines the population's fitness? Levien et al. have addressed this question by constructing a mapping between the various ways of sampling lineages from a finite population \cite{levien2020}. They have shown that Equation \eqref{EL} can be extended to the finite culture by interpreting the tree distribution of generation times  as the distribution of generation times among newborn cells in a snapshot of the population. As discussed in the previous section, the two distributions are identical; however the latter has a clear interpretation in the finite culture, while the former does not exist in the finite setting. 

In the study of bacterial physiology, it is common to observe cells in microfluidic chambers which can only contain a few cells, the most common such device being the \emph{mother-machine} \cite{susman2018aa,robert2010}. In this device, cells are trapped in small channels with one open end, each approximately the width of a single-cell and length of 5 to 10 cells. As the cell grow and divide, daughter cells are flushed out of the open end of the channels, leaving the observer with data from individuals lineages (corresponding to the mother cells at the closed end of the channel), rather than entire genealogies.    We would like to connect these observations with population dynamics, for example, by making predictions about the fate of strain in a competition experiment based on the observed variability and heritability of phenotypic traits along a lineage.   However, thus far the mathematical frameworks we have presented require either knowledge of the entire population tree in a large culture (e.g. Equation \eqref{EL2}), or assumptions about the underlying model for growth and division (e.g. Equation \eqref{Lambda_ou}). Is there a model independent way to link lineage dynamics to fitness? This question has been considered by Levien et al. in ref. \cite{levien2020b}, where a model independent algorithm for estimating of the population growth rate from independent lineages is presented.   This algorithm leverages a large deviation principle underlying the dynamics of exponentially proliferating branching processes, which relates the long-term growth rate to the fluctuations in the number of divisions among independent lineages. However, the large deviation structure has implications for the convergence of the algorithm, and it can be shown that the convergence of the estimator is non-monotonic in the duration of the lineages. Beyond the practical implications of the large deviation structure, it has inspired connections between population dynamics and fundamental problems in statistical mechanisms, such as the fluctuations theorems derived in ref \cite{genthon2020}.


\subsection{Evolutionary dynamics}\label{sec:evo}
For some applications, it is important to consider fluctuations due to the finite population size. Such fluctuations are neglected in the derivation of Equation \eqref{logistic}, which assumes both genotypes are sufficiently abundant so that their dynamics are effectively deterministic. Finite number fluctuations become relevant in evolutionary dynamics. Here, new mutations emerge in single-cells and their fate is determined by stochastic dynamics of the initial progeny. In this case, one typically studies the  \emph{fixation probability}, or the chance that a mutant will eventually dominate the population \cite{gillespie1974}. Fixation probabilities have been well-studied for numerous variations of the Moran process and related models. In the simplest setting where the environment is kept constant and each cell divides at a fixed rate, the fixation probability of a mutant with population growth rate $\Lambda_1$ emerging in a resident population with growth rate $\Lambda_2$ is approximately proportional to the normalized growth rate. Specifically,  ${\mathbb P}[\text{genotype 1 fixates}] \propto \Lambda_1/(\Lambda_1 + \Lambda_2)$ provided $\Lambda_1> \Lambda_2$ and ${\mathbb P}[\text{genotype 1 fixates}] = 0$ otherwise. The expression, which was originally obtained by Haldane \cite{haldane1927}, holds when the total population size is large. How do the complexities of real evolutionary dynamics, which occur in changing environments and involve genotypes which may express many phenotypes, affect the ability of mutants to fixate?  Some progress towards answering this question has recently been made: In ref. \cite{cvijovi2015} Cvijovic et al. studied fixation probabilities of phenotypically homogenous populations in changing environments, showing that mutations which are deleterious in the long term (meaning $\Lambda_1 < \Lambda_2$)  can have a positive fixation probability.   In ref \cite{carja2019}, it has been demonstrated that mutations which confer the ability to switch phenotypes may have a more difficult time fixating than would be expected from the classical theory. While previous works have considered how environmental and phenotypic dynamics interact to shape the fixation probabilities, very little is known about the long term evolutionary trajectories which involve sequences of many mutations fixating successively. In particular, the question of whether evolution will eventually ``find" the optimal strategy for surviving in a changing environment remains largely unanswered.

%
%

\subsection{Disentangling intrinsic and environmental variability}
In the previous sections, we implicitly assume that cell-to-cell differences are generated by  \emph{intrinsic} factors within the cells, such as stochastic fluctuations in gene expression. 
However, cell-to-cell variability is often triggered by environmental factors. While the failure of many pharmaceutical interventions in both microbial infections and in cancer is commonly attributed to phenotypic heterogeneity \cite{altschuler2010}, a detailed understanding of the interplay between intrinsic and environmental  phenotypic variability is lacking in many systems.   Without detailed information about the external environment, how does one determine what combination of environmental and intrinsic factors are responsible for the distribution of phenotypes in a population?   Van Vliet et al. have approached this problem in ref. \cite{vanvliet2018}. They tracked \emph{E. coli} lineages in a microfluidic chamber,  showing that spatial correlations in gene expression emerge from a combination of shared ancestry, spatial gradients in the environment and cell-to-cell interactions. Their approach is to consider the conditional correlations. For example, by looking at correlations of gene expression between neighboring cells which do not share a common ancestor, they can infer that such correlations are due to spatial factors or cell-to-cell interactions.

While studies such as ref. \cite{vanvliet2018} can give some insight into how various forms of phenotypic variation emerge,  the complexity of natural environments dwarfs those of any experimental setup.  For example, the causative agent of Tuberculosis, the bacteria \emph{Mycobacteria tuberculosis}, is consumed by macrophages within the host's lungs \cite{aldridge2012}. As the infection grows, a single lineage of \emph{M. tuberculosis} may grow in numerous different macrophages, which themselves move throughout the host experiencing a range of environmental conditions. At the same time, number of \emph{in vitro} studies have shown that \emph{M. tuberculosis} division is intrinsically asymmetric, producing  populations with heterogeneous growth rates, sizes and antibiotic susceptibility \cite{rego2017}.  It remains unclear how relevant this intrinsic variability is to patient outcomes because we lack a quantitive understanding of the effects of environmental variability in the host \cite{aldridge2012}.  Similar  challenges exist in the context of Salmonella, another pathogenic bacteria that propagates in host macrophages \cite{claudi2014}.  

Experimental studies which attempt to observe single-cell dynamics in more realistic environments, such as ref.  \cite{claudi2014}, are essential in order to build of more complete understanding of phenotypic variability in natural environments, but should be guided by a solid theoretical framework.  The role of theory in this context is twofold: First, it can help guide experiments toward the relevant phenotypic traits, as we have already discussed how the choice of phenotypic traits is essential; variability can be beneficial in one trait (e.g. generation times) and deleterious in another (e.g growth rates). Second, theory is essential in order to design experiments where the relevant measures of variability can actually be obtained. Some questions may require data obtained from entire population trees, while for others snapshots of the population at various time points are sufficient. 

%

 \section{Summary }
 Experimentalists now have the ability to observe single cells for hundreds of generations in highly controlled environments. Combined with the theoretical tools from populations dynamics, this data has potential to advance our understanding of the interplay between microbial physiology and evolution. This review has focused on understanding how non-genetic variation in populations affects population growth, which we view as a key step towards the more general problem of understanding how selection acts on physiological traits. Outside of the microbial context, much of the work in this area has focused on the idea that organisms evolved to diversify their phenotypes as a means to alleviate risk. The literature on this topic dates back to the seminal work of Kelly on gambling. As we showed in section \ref{sec:kelly}, Kelly derived the gambling strategy which optimizes long-term profits in the simple context where the gambler bets a fraction of their money on each trial in a sequence of independent trials \cite{kelly1956}. Long-term profits are optimized by a trade-off between maximizing the expected profits from each trial and reducing the risk of losing all of one's winnings. This idea has been shown to elucidate the fitness benefits of phenotypic switching in microbes, albeit with some technical caveats which arise when considering models where organisms grow and divide continuously; see section \ref{sec:switch}.  Kussell et al. showed that organisms face a tradeoff between maximizing their growth in each environment and maintaining the ability to quickly adapt to a  new environment \cite{kussell2005b,kussell2005}. Whether evolution actually has the potential to find the phenotypic switching rates which maximize the population growth rate remains an open question, which we discussed in section \ref{sec:evo}.

The majority of the literature on bet-hedging is focused on coarse-grained models of phenotypic variability in which the mechanisms of cell growth and division are not modeled explicitly. \cite{xue2017,starrfelt2012,childs2010,kussell2005b,kussell2005,skanata2016}. However, research has shown that these details have important implications for the population level effects of phenotypic variability.  Focusing on populations growing exponentially in constant environments, the recent studies discussed in section \ref{sec:growthratevar} have left us with diverging ideas about how this type of variability affects populations dynamics and fitness. For example, depending on the specific model one adopts for the single-cell growth rate dynamics, fluctuations in growth rates can either benefit the population or be detrimental to population fitness.  When growth rate variability is generated by asymmetric partitioning of proteins which affect growth, the effects of asymmetry on population growth depend heavily on the functional relationship between fitness and the protein concentrations; namely, whether the dependence is convex or concave. In both the case of growth rate fluctuations and asymmetric segregation, the most informative model is likely context specific.    Despite a vast body of theoretical research, we still know relatively little about how different sources of phenotypic variability interact to shape populations dynamics, as most theoretical studies have focused on simple systems which isolate a certain source of cell-to-cell variation. The next step is to build a more unified understanding of the role of phenotypic variability in populations dynamics,  which demands experiments that are informed by existing models, as well as novel models inspired by these experiments. Thus, the field presents an abundance of opportunities for biophysicists, applied mathematicians, data scientists and experimental biologists.  
 
 \section*{Acknowledgements }
We thank Bree Aldridge, Nathalie Balaban, Alvaro Sanchez and Hanna Salman for providing helpful feedback on the manuscript. We acknowledge funding support from NSF grant DMS-1902895 (EL), NSF grants DMR-1610737 and MRSEC-1420382, the Simons Foundation (JK), NSF CAREER  Award 1752024 and Harvard Dean's Competitive Fund for Promising Scholarship  (AA).

\newpage
\bibliography{variability_review.bib}

\end{document}